\documentclass[prb,twocolumn,showpacs,preprintnumbers,amsmath,amssymb]{revtex4}

\usepackage{graphicx}
\usepackage{color}
\usepackage{amsmath}
\usepackage{amssymb}
\usepackage{bm}

\begin{document}

\title[]{Temperature memory and resistive glassy behaviors of a perovskite manganite}
\author{D. N. H. \surname{Nam}}
\email{daonhnam@yahoo.com}
\author{N. V. \surname{Khien}}
\author{N. V. \surname{Dai}}
\author{L. V. \surname{Hong}}
\author{N. X. \surname{Phuc}}
\affiliation{Institute of Materials Science, VAST, 18 Hoang-Quoc-Viet, Hanoi, Vietnam}

\begin{abstract}
This paper reports the observations of long-time relaxation, aging, and temperature memory behaviors of resistance and magnetization in the ferromagnetic state of a polycrystalline La$_{0.7}$Ca$_{0.3}$Mn$_{0.925}$Ti$_{0.075}$O$_3$ compound. The observed glassy dynamics of the electrical transport appears to be magnetically originated and has a very close association with the magnetic glassiness of the sample. Phase separation and strong correlation between magnetic interactions and electronic conduction play the essential roles in producing such a resistive glassiness. We explain the observed effects in terms of a coexistence of two competing thermomagnetic processes, domain growth and magnetic freezing, and propose that hole-doped perovskite manganites can be considered as "resistive glasses".
\end{abstract}

\date[]{Received \today}
\maketitle

\section{INTRODUCTION}

Glassy systems are well known for their nonequilibrium slow dynamics such as
the long-time relaxation, aging, and memory behaviors. Although the dynamics
of electrons is generally considered very rapid, glassy transport behaviors
have been observed in various systems including electron glasses in
Anderson-insulator film structures,\cite{Vaknin3,Orlyanchik2} a 2D electron
system in Si \cite{Jaroszynski3} and mesoporous Si,\cite{Borini} ultrathin and granular metal films,\cite{Martinez-Arizala,Bielejec} rare earth
hydrides YH$_{3-\delta}$,\cite{Lee} and a number of perovskite manganite and cobaltite polycrystalline bulks,\cite{Roy,Levy,Tao,De} single crystals,\cite{Wu,Kimura,Anane} and films \cite{Helmolt1,Casa,Chen,Bhattacharya} etc.. Such a nonergodic electron dynamics is obviously not expected in conventional electronic systems and the underlying physics may not be unique considering the fact that it can be observed in a variety of materials. As for mixed-valence manganites, $R_{1-x}A_x$MnO$_3$ ($R$: trivalent are earth, $A$: divalent alkaline elements), the slow resistance relaxation and aging phenomena were noticed quite early by Helmolt \textit{et al.}\cite{Helmolt1} after the discovery of the colossal magnetoresistance (CMR) effect.\cite{Kusters,Jin} The observed glassy transport has been commonly attributed to the slow evolution of the phase conversions among competing phases coexisting in the material as a result of phase separation. Remarkably, using high-resolution transmission electron microscopy technique, Tao \textit{et al.}\cite{Tao} observed a dynamic competition between charge-ordered and charge-disordered phases in La$_{0.23}$Ca$_{0.77}$MnO$_3$ and related it to the change in the time dependence of the resistivity. However, the authors also suggested that the resistivity relaxation could be a property of the charge-ordered (orthorhombic) phase itself and not related to the phase coexistence. Kimura \textit{et al.}\cite{Kimura} compared the transport and magnetic behaviors of Cr-doped Nd$_{1/2}$Ca$_{1/2}$MnO$_3$ with the dielectric phenomena in "relaxor ferroelectrics" and proposed that the system can be viewed as a "relaxor ferromagnet". In addition, Wu \textit{et al.}\cite{Wu} observed glassy transport in La$_{1-x}$Sr$_x$CoO$_3$ cobaltites and specifically ascribed the phenomenon to the dynamics of the sample's spin-glass component.

We report here the observations of long-time relaxation, aging, and memory behaviors of resistivity, which apparently very much resemble the characteristics of spin glasses, in the ferromagnetic (FM) state of a polycrystalline La$_{0.7}$Ca$_{0.3}$Mn$_{0.925}$Ti$_{0.075}$O$_3$ compound. The resistive glassy dynamics exists in zero magnetic field but is found to be magnetically originated. Strikingly, temperature memory behaviors of both resistivity and magnetization are revealed in the ferromagnetic state. We explain the results in terms of two competitive thermomagnetic processes: domain growth and magnetic freezing. We propose that glassy transport is a common behavior of mixed-valence perovskite compounds where phase separation occurs spontaneously and the interplay between magnetism and electrical property is strong; the materials therefore can be considered as "resistive glasses".

\section{EXPERIMENTS}

The La$_{0.7}$Ca$_{0.3}$Mn$_{0.925}$Ti$_{0.075}$O$_3$ compound was prepared by a conventional solid-state reaction method. Pure ($\geq$99.99\%) raw powders with appropriate amounts of La$_2$O$_3$, CaCO$_3$, MnO$_2$, and TiO$_2$ were thoroughly ground, mixed, pelletized and then calcined at several processing steps with increasing temperatures from 900 $^\mathrm{o}$C to 1200 $^\mathrm{o}$C and intermediate grindings and pelletizations. The product was then sintered at 1300 $^\mathrm{o}$C for 48 h in ambient atmosphere. The final sample was obtained after a very slow cooling process from the sintering to room temperature with an annealing step at 700 $^\mathrm{o}$C for 5 hours. Room-temperature x-ray diffraction measurements showed that the sample is single phase of a perovskite orthorhombic (space group $Pnma$) structure. Field-cooled (FC) and zero-field-cooled (ZFC) magnetization and four-probe resistivity measurements were carried out in a Quantum Design PPMS-6000. The sample used for resistivity measurements was a 1.9$\times$1.2$\times$7.5 $\mathrm{mm}^3$ rectangular bar that was firmly glued to the PPMS puck to have a good thermal contact but electrically separated from the puck by a sheet of cigarette paper. To reduce Joule heating and current induced effects, a very small dc current $I=100$ nA (corresponding to a current density of 4.4 $\mu$A/cm$^2$) driven in AC mode was used. A low heating and cooling rate of 1 K/min was always chosen for all of the measurements.

\section{RESULTS AND DISCUSSION}

\subsection{Magnetic and transport characterizations}

\begin{figure}[t!]
\includegraphics[width=7.5cm]{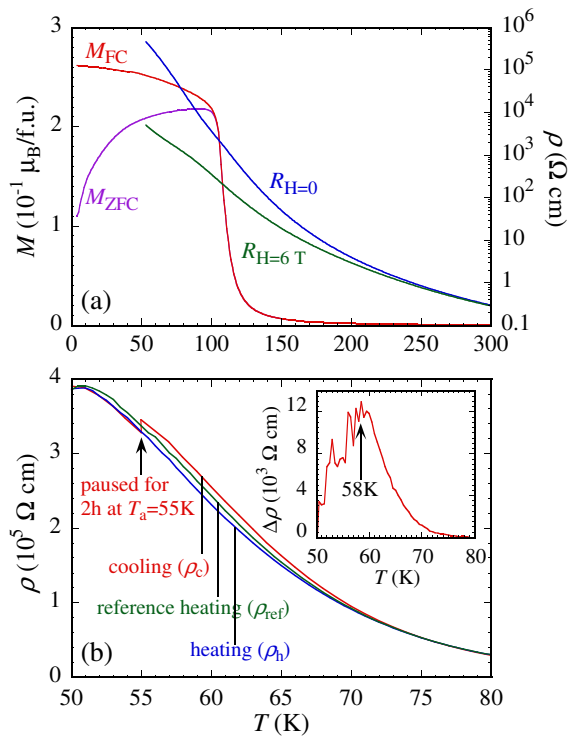}
\caption{(color online). (a) $M_\mathrm{ZFC}(T)$, $M_\mathrm{FC}(T)$ (left axis, $H=100$ Oe) and $\rho(T)$ (right axis, $H=0$ and 6 T) of La$_{0.7}$Ca$_{0.3}$Mn$_{0.925}$Ti$_{0.075}$O$_3$. (b) $\rho(T)$ measured by different protocols: $\rho_\mathrm{ref}(T)$ was measured on heating following a continuous cooling from 300 K to 50 K, $\rho_\mathrm{c}(T)$ was recorded during cooling the sample to 50 K with a pause at 55 K for 2 h and $\rho_\mathrm{h}(T)$  was then measured on reheating. The inset of (b) shows the reduced resistivity $\Delta\rho(T)=\rho_\mathrm{ref}(T)-\rho_\mathrm{h}(T)$ that exhibits a maximum at $\sim$58 K.} \label{fig1}
\end{figure}

According to the double-exchange (DE) mechanism for mixed-valence manganites,\cite{Zener} the transfer of an $e_\mathrm{g}$ electron between two adjacent Mn ions occurs only when their localized $t_\mathrm{2g}$ magnetic moments are aligned in parallel, implying that metallic conducting behavior should come along with a FM state. Nevertheless, manganites are strongly separated systems, so that the ideal DE correlation between magnetism and electrical transport may not be always obtained. While La$_{0.7}$Ca$_{0.3}$MnO$_3$ is a typical DE ferromagnet that has a metallic conducting behavior in the FM phase below $T_\mathrm{c}\approx250$ K, since Ti$^{4+}$ ions do not take part in magnetic couplings, substitution of Mn by Ti destabilizes the network of Mn ions, leading to a deterioration of both DE ferromagnetism and conductivity.\cite{Liu,Nam4} Temperature dependent magnetizations, $M_\mathrm{ZFC}(T)$ and $M_\mathrm{FC}(T)$, and resistivity, $\rho(T)$, in Fig. 1(a) show that the La$_{0.7}$Ca$_{0.3}$Mn$_{0.925}$Ti$_{0.075}$O$_3$ compound is an insulator ferromagnet with $T_\mathrm{c}\approx108$ K. The insulating behavior in the FM state would suggest that the sample is segregated into metallic conducting FM regions separated by non-FM insulating boundaries. With decreasing temperature, the resistance increases continuously and goes beyond the limit of our measurement system below 50 K. Under the influence of a magnetic field $H=6$ T, the transport behavior remains that of an insulator but the resistance is strongly suppressed leading to a negative magnetoresistance, $-MR=(R_{H=0}-R_{H=6\mathrm{T}})/R_{H=\mathrm{6T}}$, which increases with lowering temperature and reaches up to $\sim$9000\% at 50 K. The negative magnetoresistance is probably due to an expansion of the DE FM metallic conducting regions that is more favored at lower temperatures.

\subsection{Temperature memory and resistive glassy behavior}

The $\rho(T)$ data in Fig. 1(a) were measured on reheating the sample from 50 K following a continuous cooling from 300 K. The $\rho(T)$ curve in zero field is replotted in Fig. 1(b) as a reference heating curve, denoted as
$\rho_\mathrm{ref}(T)$. In other measurements, instead of cooling down
the sample continuously directly to 50 K, the cooling was paused at a constant
temperature $T_\mathrm{a}<T_\mathrm{c}$ for a duration time $t_\mathrm{a}$.
Interestingly, we observed a clear resistance relaxation that had no
tendency to stop even after several hours during the pause, showing a signature of an aging process. The cooling was then resumed to cool the sample to 50 K. A typical $\rho(T)$ curve recorded during this cooling procedure with $T_\mathrm{a}=55$ K and $t_\mathrm{a}=2$ h are plotted as $\rho_\mathrm{c}(T)$ in Fig. 1b, which shows a step at $T_\mathrm{a}$ caused by the aging. Right after reaching 50 K, the temperature was swept back and $\rho_\mathrm{h}(T)$ data were collected. Intuitively, curves $\rho_\mathrm{ref}(T)$ and $\rho_\mathrm{h}(T)$ are not the same, indicating a cooling history dependence. By subtracting $\rho_\mathrm{h}(T)$ from $\rho_\mathrm{ref}(T)$ we obtained a temperature dependence of the reduced resistivity, $\Delta\rho(T)$, that shows a maximum at $\sim$58 K [the inset of Fig. 1(b)], which is very close to the temperature where the pause was made. This feature seems to imply that the pause at $T_\mathrm{a}$ during cooling was imprinted into the system and can be retrieved on reheating, signaling the presence of a temperature memory effect that was previously observed in magnetic glassy systems.\cite{Jonason,Jonsson,Mathieu1} Henceforth, we refer the measurement of a heating curve following a paused cooling as "memory measurement" (or "memory curve"). In the case where one pause is made, the memory characterized by a maximum of $\Delta\rho(T)$ is then called "single temperature memory" (SME) and $\Delta\rho$ is denoted as $\Delta\rho_\mathrm{SME}^{T_\mathrm{a}}$.

\begin{figure}[t!]
\includegraphics[width=7.5cm]{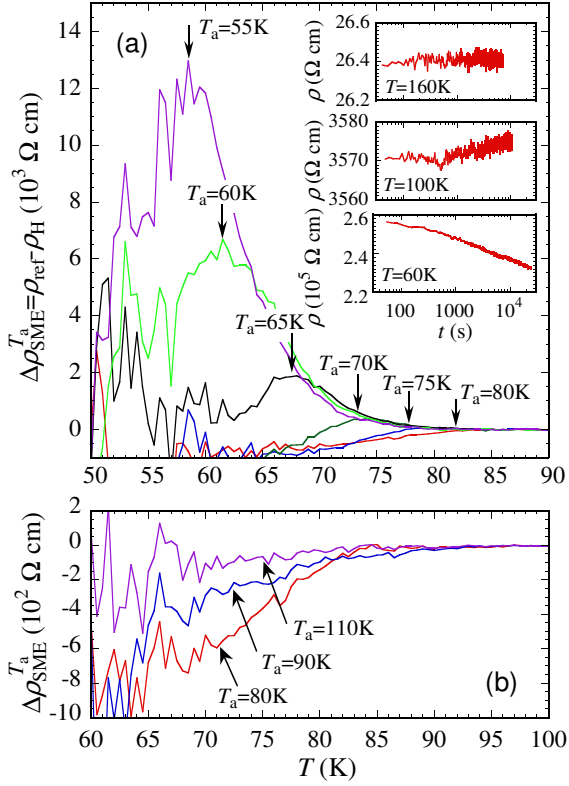}
\caption{(color online). Single temperature memory curves,
$\Delta\rho_\mathrm{SME}^{T_\mathrm{a}}(T)$, with different $T_\mathrm{a}$'s. (a) $\Delta\rho_\mathrm{SME}^{T_\mathrm{a}}(T)$ curves for $T_\mathrm{a}=55$, 60, 65, 70, 75, and 80 K. The arrows mark the
$\Delta\rho_\mathrm{SME}^{T_\mathrm{a}}(T)$ maximum with the corresponding
$T_\mathrm{a}$. The
insets display $\rho(t)$ curves measured at
$T_\mathrm{a}=60$, 100, and
160 K. (b) $\Delta\rho_\mathrm{SME}^{T_\mathrm{a}}(T)$ for
$T_\mathrm{a}=80$, 90, and 110 K. The arrows mark the
$\Delta\rho_\mathrm{SME}^{T_\mathrm{a}}(T)$ curve with the corresponding
$T_\mathrm{a}$.} \label{fig2}
\end{figure}

Observations of magnetic glassy behaviors similar to those of canonical spin
glasses in DE perovskite cobaltites and manganites have been reported in a number of publications.\cite{Nam1,Nam2,Mathieu2,Nam5} However, although signatures of glassy transport such as slow relaxation and aging were sometimes observed, memory behaviors of resistance have not been explored so far. To confirm the memory behavior in Fig. 1(b), we carried out a series of similar experiments with $t_\mathrm{a}=2$ h but various temperatures $T_\mathrm{a}$'s; the results are presented in Fig. 2. The maximum of $\Delta\rho_\mathrm{SME}^{T_\mathrm{a}}$ near $T_\mathrm{a}$ is reproducibly obtained but strongly suppressed with increasing $T_\mathrm{a}$ and finally vanished for $T_\mathrm{a}>T_\mathrm{c}$; i.e., it is well defined for $T_\mathrm{a}=55$, 60, 65, 70, and 75 K, but appears less and less defined with further increasing $T_\mathrm{a}$ and becomes unresolvable for  $T_\mathrm{a}=110$ K. Irrespective of $T_\mathrm{a}$, $\Delta\rho_\mathrm{SME}^{T_\mathrm{a}}$ is strongly suppressed with temperature and basically becomes zero in the paramagnetic state. Although a magnetic field is not required for the observation of this memory effect, these results unambiguously indicate that it is indeed magnetically originated. Another noticeable feature is the tendency of $\Delta\rho_\mathrm{SME}^{T_\mathrm{a}}$ at $T<T_a$ to change its sign to
negative when $T_\mathrm{a}$ approaches close to $T_\mathrm{c}$. Resistivity
relaxation curves, $\rho(t)$, measured during the pauses at $T_\mathrm{a}=60$ K ($\ll T_\mathrm{c}$), 100 K ($\approx T_\mathrm{c}$), and 160 K ($\gg T_\mathrm{c}$) plotted in the insets of Fig. 2(a) clearly show a very
strong downward relaxation at 60 K, a weaker upward relaxation at 100 K, and no relaxation at all at 160 K. Qualitatively, the changes in both amplitude and
sign of $\rho(t)$ are consistent with the variation of the
$\Delta\rho_\mathrm{SME}^{T_\mathrm{a}}$ curves. Although the underlying
physics is probably different, this behavior is reminiscent of the change in
relaxation direction of the field-cooled magnetization in the vicinity of the glass phase transition in spin glasses and random magnets.\cite{Jonsson,Nam5}

\begin{figure}[t!]
\includegraphics[width=7.5cm]{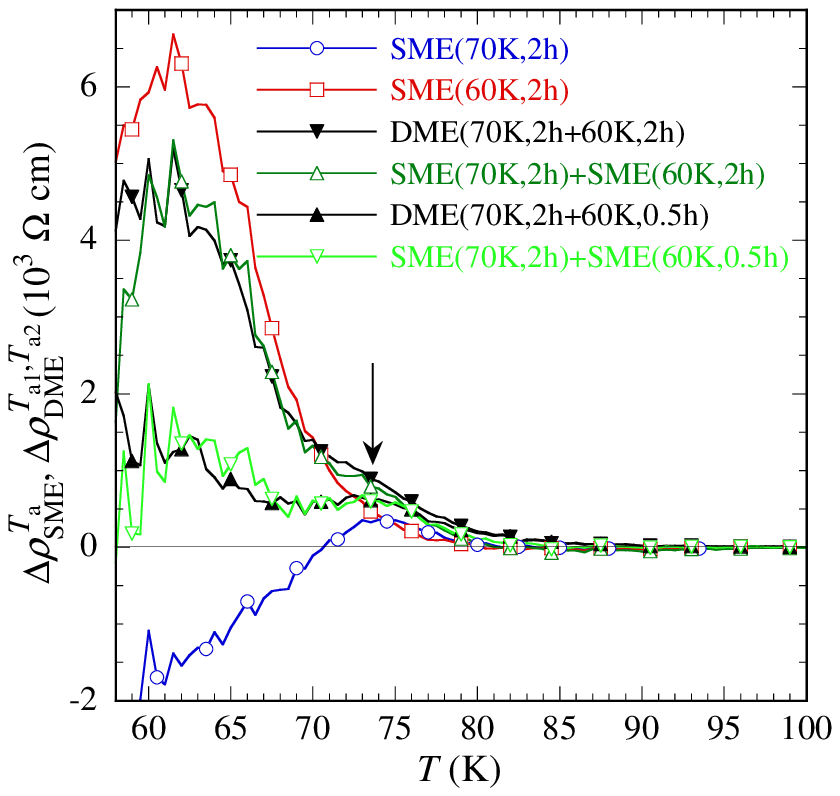}
\caption{(color online). Double temperature memory as a combination of the corresponding single
temperature memories,
$\Delta\rho_\mathrm{DME}^{T_\mathrm{a1},T_\mathrm{a2}}(T)=\Delta\rho_\mathrm{SME}^{T_\mathrm{a1}}(T)+\Delta\rho_\mathrm{SME}^{T_\mathrm{a2}}(T)$.
The $\Delta\rho_\mathrm{DME}^{T_\mathrm{a1},T_\mathrm{a2}}(T)$ curve shows a
maximum near $T_\mathrm{a2}=60$ K and a shoulder (marked by an arrow) near
$T_\mathrm{a1}=70$ K. $t_\mathrm{a1}=2$
h. Symbols: for $t_\mathrm{a2}=2$ h:
$\circ=\Delta\rho_\mathrm{SME}^\mathrm{70 K}$,
$\Box=\Delta\rho_\mathrm{SME}^\mathrm{60 K}$,
$\blacktriangledown=\Delta\rho_\mathrm{DME}^\mathrm{70K,60K}$,
$\vartriangle=\Delta\rho_\mathrm{SME}^\mathrm{70
K}+\Delta\rho_\mathrm{SME}^\mathrm{60 K}$; for $t_\mathrm{a2}=0.5$ h:
$\blacktriangle=\Delta\rho_\mathrm{DME}^\mathrm{70K,60K}$,
$\triangledown=\Delta\rho_\mathrm{SME}^\mathrm{70
K}+\Delta\rho_\mathrm{SME}^\mathrm{60 K}$.} \label{fig3}
\end{figure}

The results in Fig. 2 clearly demonstrate that the system memorizes the
temperature where an event was made during cooling and recalls it on reheating. To figure out whether the system can memorize separately more than one event in a single cooling (thus yielding a "multiple temperature memory"), we
made two pauses with the same $t_\mathrm{a}=2$ h sequently at
$T_\mathrm{a1}=70$ K and $T_\mathrm{a2}=60$ K in the same cooling process. The
corresponding $\Delta\rho$ curve (Fig. 3) shows a large maximum near 60 K and a small shoulder around 70 K, indicating a "double temperature memory" (DME)
behavior. Moreover, this DME is in fact a combination of the two SME's at the
corresponding $T_\mathrm{a}$'s; i.e., the summation of the two SME curves with
$T_\mathrm{a}=60$ K and 70 K, $\Delta\rho_\mathrm{SME}^\mathrm{60K}(T)$ and
$\Delta\rho_\mathrm{SME}^\mathrm{70K}(T)$ respectively, matches very well the
DME curve $\Delta\rho_\mathrm{DME}^\mathrm{70K,60K}(T)$, as illustrated in Fig. 3. Just for another example, we reduced $t_\mathrm{a}$ of the pause at 60 K to
0.5 h in both the SME and DME experiments and again observed that the superposition of two SME's gives the corresponding DME. Unambiguously, the system can memorize multiple events separately created on a single cooling and store them without interference -- a behavior that has been well observed in magnetic glasses.\cite{Jonason,Jonsson,Nam2,Mathieu2}

The use of temperature as the only parameter to vary the dynamics of the system allows probing the system by its original dynamics. The observed nonequilibrium behaviors of resistance bear striking resemblances with the dynamic characteristics of magnetic glasses. The resistance keeps relaxing towards equilibrium in such a long period of time that could go beyond the laboratory time scale even in a constantly stabilized condition. After being "aged" at the pauses during cooling, the system responds differently on reheating, demonstrating a clear age-dependent behavior. Furthermore, the temperature memory effect, an inherent characteristics of magnetic glasses, is also observed. Since the resistive glassy behaviors are observed only in the FM state, they must have a magnetic origin. The insulating behavior in FM state suggests that the sample is not a pure DE ferromagnet, but probably a separated system consisting of two competing phases: DE FM metallic-conducting domains separated by a non-FM insulating matrix. Low temperatures or high magnetic fields both favor the DE interaction and expand the FM domains at the
expense of the non-FM insulating matrix. During the pause below
$T_\mathrm{c}$, the FM domains would evolve with time by polarizing the spins surrounding to expand the conducting regions. On the other hand, thermal activation helps the domain moments relax towards directions which are energetically favored by their randomly distributed local anisotropy,
therefore increasing the spin scattering of electrons traveling between domains. The domain growth seems to be dominant at low temperatures, resulting in a downward resistivity relaxation (Fig. 2). In contrast, magnetic moments relax faster at higher temperatures, overcoming the domain growth process at temperatures close to $T_\mathrm{c}$ and therefore producing an upward resistivity relaxation.

We adopt the N\'{e}el's relaxation law
\cite{Neel} to qualitatively describe the freezing of FM domains,
$\tau=\tau_0\exp(E_\mathrm{a}/k_\mathrm{B}T)$, where $E_\mathrm{a}$ is the
anisotropy energy that is proportional to the domain size, $k_\mathrm{B}$ is
the Boltzmann's constant, and $\tau$ and $\tau_0$ are the relaxation and microscopic relaxation times, respectively. Lowering temperature leads to an exponential increase of $\tau$, causing the domains to become deeper frozen. One of the key ingredients of the temperature memory effect is the freezing of magnetic moments in random directions that preserves the "aged" magnetic configuration when cooling is resumed after the pause. The frozen moments would still relax but with very much smaller rates at lower temperatures. Upon reheating, the melting of the aged configuration returns the system to the state without aging above $T_\mathrm{a}$, exhibiting the observed memory effect. Since after the pause the magnetic moments relaxed into directions with higher energy barriers [$k_\mathrm{B}T_\mathrm{a}\ln(t_\mathrm{a}/\tau_0)$, according to the N\'{e}el's law] than $k_\mathrm{B}T_\mathrm{a}$, higher temperatures are then required to remove them from the freezing directions. This explains why the maximum of $\Delta\rho(T)$ is always at a higher temperature than $T_\mathrm{a}$. In a real system, there always exist wide distributions of $\tau_0$ and $E_\mathrm{a}$. Magnetic moments of domains with $E_\mathrm{a}\leq k_\mathrm{B}T_\mathrm{a}$ rotate freely under the thermal excitation, those with $E_\mathrm{a}\gg k_\mathrm{B}T_\mathrm{a}$ are effectively frozen, and only those with $E_\mathrm{a}\gtrsim k_\mathrm{B}T_\mathrm{a}$ contribute to the relaxation at $T_\mathrm{a}$. This explains why we could observe a multiple temperature memory. Ideally, if $T_\mathrm{a2}\ln(t_\mathrm{a2}/\tau_0)<T_\mathrm{a1}$, there could be no
interference between the two aged configurations. $\Delta\rho(T)$ approaches
zero only at sufficiently high temperature far above $T_\mathrm{a}$ where there is effectively no difference in domain sizes between the memory and the
reference configurations.

\begin{figure}[t!]
\includegraphics[width=7.5cm]{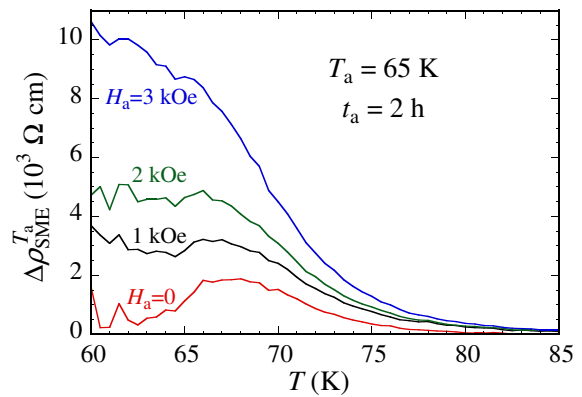}
\caption{The influence of magnetic field on the nonequilibrium resistance
dynamics. A magnetic field $H_\mathrm{a}$ is turned on during the
pause at $T_\mathrm{a}=65$ K on cooling. The application of $H_\mathrm{a}$ causes a "crosstalk" between the memory established at $T_\mathrm{a}$ and the domain configurations developed during further cooling. The memory maximum in the $\Delta\rho_\mathrm{SME}^{65K}(T)$ curve is unresolvable at
$H_\mathrm{a}=3$ kOe.} \label{fig4}
\end{figure}

In order to examine the influence of magnetic field on the memory effect, a
magnetic field $H_\mathrm{a}$ was turned on during the pause. Right after the
aging time $t_\mathrm{a}$, $H_\mathrm{a}$ was removed and the cooling was
immediately resumed. The SME $\Delta\rho(T)$ curves measured with different
$H_\mathrm{a}$'s are presented in Fig. 4. The field is expected to favor the
domain growth and therefore to increase the $\Delta\rho(T)$ maximum near
$T_\mathrm{a}$. However, the freezing of magnetic moments are also strongly
affected. Instead of relaxing towards random directions, the magnetic moments
are polarized and relax preferably towards the external field direction,
establishing a configuration that is advantageous for the unfrozen domains to
grow such that the domain growth during further cooling could effectively
overwhelm the aged configuration previously established at $T_\mathrm{a}$. We
can see clearly in Fig. 4 that, while the maximum of the $\Delta\rho(T)$ curves is indeed raised with increasing $H_\mathrm{a}$, it is significantly broadened and spreads towards lower temperatures. Apparently, higher $H_\mathrm{a}$ applied at $T_\mathrm{a}$ causes a stronger development of magnetic domains upon further cooling, which may interfere with the aged configuration pre-established at $T_\mathrm{a}$. With $H_\mathrm{a}=3$ kOe, the maximum of $\Delta\rho(T)$ is no longer defined, suggesting that the development of magnetic domains with decreasing temperature below $T_\mathrm{a}$ becomes entirely dominant.

\begin{figure}[t!]
\includegraphics[width=7.5cm]{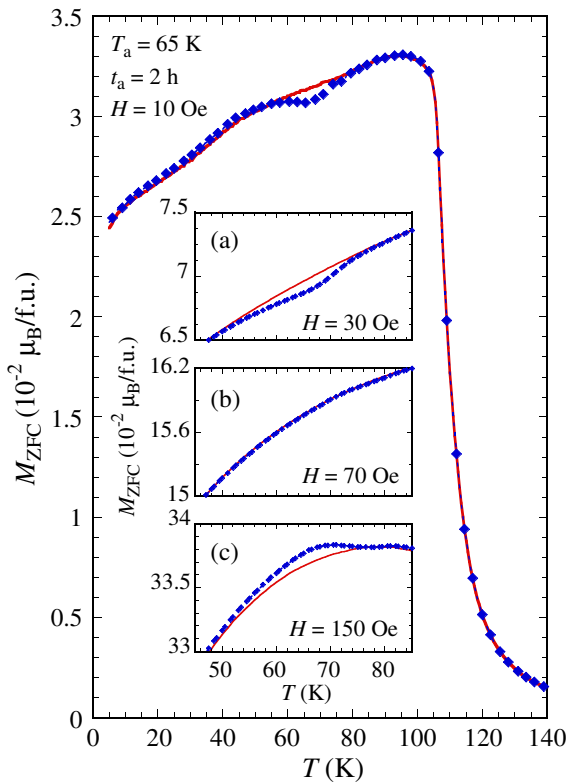}
\caption{(color online). The temperature memory behavior observed on $M_\mathrm{ZFC}(T)$. For the memory curves (symbols), the sample was zero-field-cooled from 300 K to 5 K with a pause at $T_\mathrm{a}=65$ K for $t_\mathrm{a}=2$ h. For the reference curves (lines), the sample was cooled directly to 5 K. After the temperature had reached 5 K, a probing field $H$ was immediately applied and $M_\mathrm{ZFC}(T)$ was recorded on reheating. Main figure: $H=10$ Oe, insets (a), (b), and (c): $H=30$, 70, and 150 Oe, respectively. Sufficiently high probing fields can align the moments of the aged domains toward its direction, turning the memory "dip" lying below into a "cusp" lying above the reference
curve.} \label{fig5}
\end{figure}

\subsection{Temperature memory behavior of magnetization}

Since the memory behavior of resistance is the result of thermomagnetic dynamic processes, it is natural to expect that similar effects can be observed in conventional magnetic measurements, even though the sample is in a ferromagnetic state. The results in Fig. 5 verify the temperature memory behavior of magnetization. With similar cooling procedures as those used for the transport measurements, depending on the probing field strength, the SME $M_\mathrm{ZFC}(T)$ curve exhibits a dip or a cusp around $T_\mathrm{a}$ (Fig. 5 insets). For very small fields, the $M_\mathrm{ZFC}$ at a given temperature is contributed from FM domains having energy barriers lower than $k_\mathrm{B}T$ that can rotate toward the field direction. The freezing of magnetic moments that relaxed into higher energy barriers at $T_\mathrm{a}$ during the pause on cooling is thus reflected by a dip in the memory curve lying below the reference curve. Melting these moments with increasing $T$ above $T_\mathrm{a}$ drives the memory magnetization toward the reference one. The effect of larger domain sizes achieved by the pause is not exhibited because their moments are frozen in random directions. However, for high probing fields, the rotation of frozen moments towards the field direction becomes significant; the larger FM domain sizes in the memory configuration (than in the corresponding reference one) give extra contributions to the magnetization. As a result, with increasing probing field, the memory dip is gradually shallowed and develops into a cusp lying above the reference curve . This influence of magnetic field on the memory behavior is thus in agreement with the picture of coexisting domain growth and magnetic freezing.

\section{CONCLUSION}

Nonequilibrium dynamic behaviors such as long-time relaxation, aging, and temperature memory of both resistivity and magnetization have been revealed in the insulator ferromagnet La$_{0.7}$Ca$_{0.3}$Mn$_{0.925}$Ti$_{0.075}$O$_3$. The glassy transport has a very close association with the magnetic glassiness and appears to be magnetically originated. These results can be qualitatively explained in terms of two competing thermodynamic processes: domain growth and magnetic freezing. The results of our dynamical study evidence the presence of phase separation and strong correlation between magnetism and electrical transport in the CMR perovskite manganite. We propose that resistive glassiness is a common character of mixed-valence perovskite manganites as a direct result of both electronic and magnetic phase separation. This nonequilibrium dynamics may be found profound in systems where there exists a strong competition between conducting and insulating phases, most likely near the metal-insulator transition with doping. Since the temperature memory behavior of resistance is associated with the freezing of magnetic domains, it is expected to be observable in the manganite insulators where FM metallic domains are embedded in the non-FM insulating matrix.

\begin{acknowledgments}
This work has been performed using facilities of the State Key Labs (IMS,
VAST).
\end{acknowledgments}

\end{document}